\newcommand{\n}{\hat{\bf n}}		

\documentstyle[multicol,aps,psfig]{revtex}
\begin{document}

\draft


\title{Anomalous force diffusion in nearly--ordered
packings of frictionless discs}

\author{D.A. Head\dag}

\address{\dag Department of Physics and Astronomy,
Vrije Universiteit 1081 HV, Amsterdam, The Netherlands}

\maketitle

\begin{abstract}
We derive analytic expressions for force propagation
in packings of frictionless discs with a narrow distribution
of disc sizes, by expanding to first order about the
known ordered solution.
The distribution of contact forces $P(f)$ is found to
be narrow at the upper surface, and broaden at
a rate that varies with depth, being superdiffusive
near the surface until crossing over to a subdiffusive
regime near the fixed base.
Furthermore, the response to an isolated load
propagates along the edge of a `cone,' as in the
ordered case, but fluctuates under ensemble
averaging by an amount that depends purely
on height, not on the lateral position.
Finally, we comment on ways in which the analytical
framework presented here could be extended to a
wider range of granular packings.
\end{abstract}



\begin{multicols}{2}

\section{Introduction}
\label{s:introduction}

The mechanical properties of static bead packings
remains a  controversial and largely unanswered
problem of granular
media~\cite{proc,jpb_proceedings,doubleY3,guyon}.
Experiments and numerical simulations have thus far
been unable to provide a definitive description of the
propagation of stresses through rigid bead packings,
in part because of the limited range of bead properties
and system sizes typically considered.
Theory is not yet at a sufficiently advanced stage to
resolve these deficiencies.
First--principle approaches such as those initiated
in~\cite{loop_forces,cambridge,gay_silveira1,gay_silveira2}
may eventually generate some form of `complete' solution,
but their significant complexity has thus far ruled
out easily verifiable predictions.
Analytically tractable theories have been devised,
such as the $q$--model~\cite{q_model,jaco_1,jaco_2}
and the force ray splitting `double Y'
model~\cite{doubleY3,doubleY1,doubleY2},
but only after making {\em ad hoc} if intuitive
assumptions regarding the local propagation of forces.

An alternative approach to treating granular piles
within an analytical framework is to restrict attention to
a specific class of packings, but otherwise treat it
as exactly as possible.
Frictionless spheres are an ideal test case:
the smooth surfaces cannot
support tangential contact forces,
removing indeterminacy problems that arise
with frictional surfaces,
and furthermore the spherical nature of the beads
means that the remaining normal component must
pass through the bead centre, so that the torque
applied at each contact must vanish.
Nonetheless disordered frictionless sphere packings
are far from trivial:
apart from mean field--type
analysis~\cite{alexei1}, results obtained
thus far have typically been numerical,
either for small systems with arbitrary
disorder~\cite{robust,adap_net},
or much larger systems within the
`approximation of small displacements'
of Roux {\em et al.}~\cite{roux_pre,roux_epl}
(in this context, it is perhaps also relevant to mention
the large--scale simulations of Breton {\em et al.}
on geometrically ordered packings with disorder
in the contact friction~\cite{breton}).

For a completely ordered lattice of frictionless discs,
all of the forces propagate along the directions
of colinear bonds, and it is straightforward to write down
all of the contact forces in response to a specified loading.
In this paper, we expand about this ordered solution
to first order in the deviations of bead radii from their
mean value, and derive analytical expressions for
the propagation and distribution of forces for piles
of arbitrary size.
This we are able to eliminate statistical noise.
Our findings are fully detailed in the subsequent sections,
but in brief:
the distribution of contact forces $P(f)$ is narrow near
the surface, and broadens as one moves further
down the pile.
This broadening is initially superdiffusive
({\em i.e.} faster than the square root of distance
from the surface),
but crosses over to a subdiffusive regime lower down
the pile and ceases to broaden precisely at the base.
Turning to consider the Green's function problem
of the response in contact forces to an isolated point load,
we find that mean of this response is concentrated
along a `cone' extending downwards from the point source
along lines of contacts.
Although the response inside the cone averages to zero,
the variance in the
fluctuations under ensemble averaging can be calculated,
and is found to depend purely on the height,
not (as might be expected) the distance from edge of the cone.

This paper is arranged as follows.
In Sec.~\ref{s:preliminaries} we specify in detail the system
to be considered.
The statistical properties of the geometry of the packing
under the specified construction procedure and
boundary conditions are derived in Sec.~\ref{s:main}.
The force propagation from a single loaded bead
within a particular packing is detailed in Sec.~\ref{s:point_load},
which is then averaged under two types of loading
to give the distribution of contact forces $P(f)$ in Sec.~\ref{s:force_diffusion}.
In Sec.~\ref{s:greens} the response to an isolated load
is found, including its statistical properties under ensemble
averaging.
Finally, in Sec.~\ref{s:disc} we summarise
and suggest possible ways in which the analytical
framework presented here might be extended.
We remark that our approach is similar in spirit to
Roux {\em et al.}'s `assumption of small displacements,'
although here were consider low--density packings for which
the disorder cannot make or break contacts.
In fact, it is essentially the small--polydispersity of the
adaptive network algorithm of Tkachenko and
Witten~\cite{adap_net}, treated analytically rather than
numerically.

\section{Preliminaries}
\label{s:preliminaries}


Here we specify the class of granular packings
under consideration in this paper,
as well as the reasons for their analytically tractability.
Firstly, we consider packings that are marginally
rigid or {\em isostatic}~\cite{alexei1,isostatic},
{\em i.e.} the number of (scalar) constraints imposed by
the conditions of force and torque balance
on each bead equals the number of
degrees of freedom in the contact forces.
Then the force network can be determined purely
from the requirements of local force and torque balance,
without reference to the deformation properties of the beads.
Also, since both the force and torque balance equations
are linear, the full contact network can be found by
simply summing the forces propagating from each loaded bead.

Secondly, our packings are constructed from
{\em frictionless} beads, so that only the normal components
of the contact forces are non--zero.
This means that the direction of the contact forces
can be uniquely determined from the geometry.
Furthermore we use {\em spherical} beads, so that
these contact forces point through the bead centres
and all torques trivially vanish.
It is straightforward to calculate the isostatic limit
for such packings.
If the mean number of contacts per bead is $Z$,
this represent $Z/2$ degrees of freedom in the
(normal) contact forces,
and $d$ constraints imposed by force balance
in a $d$--dimensional system.
Thus the isostatic limit for frictionless spheres
has $Z=2d$.

Thirdly, we suppose that the beads are
{\em almost monodisperse} in size, {\em i.e.} the bead radii
are narrowly distributed about some mean value.
Then the contact topology will be the same as
for a completely ordered system.
Since $Z=2d$ everywhere,
the load on any one bead can be
uniquely decomposed into the $d$ contact forces
of the two lower beads
(assuming there are no horizontal contacts).
In this manner the forces due to external loads
will propagate downwards through the system
until reaching the fixed base~\cite{adap_net,mult_noise}.

Finally, the packing is formed by the
{\em sequential deposition} of beads
to the upper surface,
and ensuring mechanical stability between each addition.
This exploits a happy coincidence of numbers:
for $Z=2d$ packing, each incoming bead will rest on $d$
others, which is also the number of contacts required
to determine the position of the added bead.
Thus the geometry of the pile can be found
without needing to know the incoming bead trajectories.

\section{Geometry of the packing}
\label{s:main}

In this section, the position vectors of the beads for
a particular set of bead radii is derived
for the $d=2$ case of frictionless discs.
An assumption underlying the analysis throughout
this paper is that all of the contact forces are compressive.
This is crucial, since granular particles are typically
regarded as non--cohesive and therefore cannot
support tensile contacts.
It is clear that such a state must exist if the
disorder is sufficiently small, since all contact forces
are compressive in the corresponding ordered limit~\cite{formal}.
The range of validity of this assumption will be discussed
in Sec.~\ref{s:force_diffusion} after the force
distribution has been found.

\label{s:geometry}

\begin{figure}
\centerline{\psfig{file=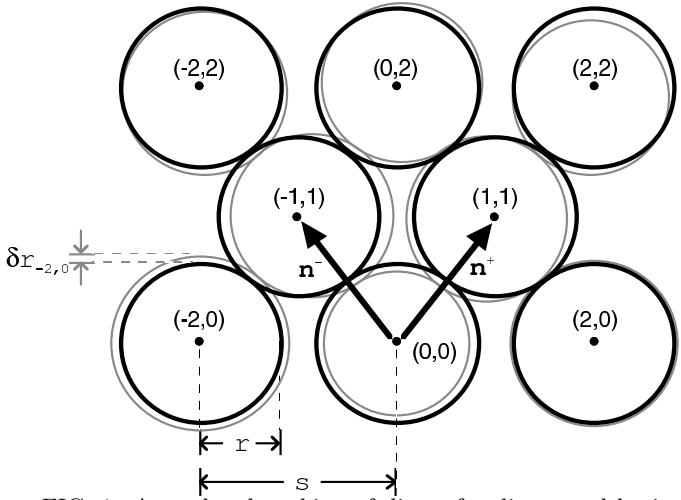,width=9cm}}
\caption{An ordered packing of discs of radius $r$
and horizontal separation $s$ (thick black circles).
The lattice indices $(i,j)$ and the lattice base vectors
$\n^{\pm}$ are also shown.
A particular example of a disordered packing,
in which the beads have radii $r+\delta r_{ij}$,
is superimposed in grey.
}
\label{f:notation}
\end{figure}

Before considering the near--crystalline packing,
it is useful to describe the completely ordered case,
both as reference and to fix the notation.
See Fig.~\ref{f:notation}.
Without disorder, each bead has radius $r$
and is located at a lattice site $(i,j)$,
where $i$ and $j$ refer to the horizontal and
vertically--upwards lattice directions, respectively.
The centre of the bead at $(i,j)$ is denoted by the
position vector ${\bf x}_{ij}$, where

\begin{equation}
{\bf x}_{ij} = r(j+i)\n^{+}
+r(j-i)\n^{-}
\end{equation}

\noindent{}Here, $\n^{\pm}$ are the unit base lattice vectors
in the upwards--right ($\n^{+}$) and upwards--left
($\n^{-}$) directions.
It is straightforward to show that
$\n^{\pm}=(\pm n_{x},n_{y})$
with $n_{x}=s/4r$, 
where $s$ is the horizontal distance between bead centres.
Each bead makes $2d=4$ contacts with beads
aligned along the lattice diagonals,
{\em i.e.} bead $(i,j)$ touches beads $(i\pm1,j\pm1)$.
Thus only sites with $i+j$ even are occupied.
The requirement that beads cannot overlap in
either the horizontal or vertical directions
fixes $s$ to be in the range $(2r,2\sqrt{3}r)$.

Disorder is introduced into the packing by
assigning each bead a radius $r+\delta r_{ij}$
prior to deposition onto lattice site $(i,j)$,
where the $\delta r_{ij}$ are random variables drawn
from a distribution
with zero mean and variance $\sigma^{2}_{\delta r}$.
When a bead is added to the growing surface,
its centre ${\bf x}_{ij}+\delta{\bf x}_{ij}$ must be
chosen to ensure the distance to both supporting
beads is equal to the sum of the corresponding radii;
to first order in the $\delta r_{ij}$\,,
this gives

\begin{eqnarray}
\delta{\bf{x}}_{ij}
&=&
\frac{1}{2}\left(\delta{\bf{x}}_{i-1j-1}
+\delta{\bf{x}}_{i+1j-1}\right)
\nonumber\\
&&-
\frac{1}{2}
\left(\begin{array}{cc}
0 & \frac{n_{y}}{n_{x}} \\
\frac{n_{x}}{n_{y}} & 0
\end{array}\right)
(\delta{\bf{x}}_{i+1j-1}-\delta{\bf{x}}_{i-1j-1})
\nonumber\\
&&+
(\delta r_{ij}+\delta r_{i-1,j-1})\,{\bf b}^{+}
+
(\delta r_{ij}+\delta r_{i+1,j-1})\,{\bf b}^{-}
\label{e:disp}
\end{eqnarray}

\noindent{}where ${\bf b}^{\pm}$ are the
reciprocal lattice vectors
${\bf b}^{\pm}=(\pm1/2n_{x},1/2n_{y})$,
{\em i.e.} vectors ${\bf b}^{\pm}$
such that $\n^{\pm}\cdot{\bf b}^{\pm}=1$
and $\n^{\pm}\cdot{\bf b}^{\mp}=0$.

We now fix the lower boundary condition to be
that the centres of all beads in the base $j=0$
have the same height and are uniformly spread horizontally:
${\bf x}_{i0}=(2irn_{x},0)$, just as in the ordered case.
Then (\ref{e:disp}) can be extrapolated to $j=0$
{\em via} an induction process to give
the perturbed bead positions purely in
terms of the $\delta r_{ij}$,

\begin{eqnarray}
\delta{\bf x}_{ij}&=&
\left\{
\delta r_{i-j,0} + \delta r_{ij}
+2\sum_{m=1}^{j-1}
\delta r_{i-(j-m),m}
\right\}
{\bf b}^{+}
\nonumber\\
&&+
\left\{
\delta r_{i+j,0}+\delta r_{ij}
+2\sum_{m=1}^{j-1}
\delta r_{i+(j-m),m}
\right\}
{\bf b}^{-}
\label{e:delta_x_ij}
\end{eqnarray}

\noindent{}Thus the position of the bead at $(i,j)$
depends on $\delta r_{ij}$,
$\delta r_{i\pm1,j-1}$,
$\delta r_{i\pm2,j-2}$ and so on,
{\em i.e.} only on the $\delta r_{ij}$ lying along
the two diagonals stretching from $(i,j)$ to the base,
hereafter referred to as the `cone.'
Note that (\ref{e:delta_x_ij})
ignores any horizontal boundaries,
and is therefore valid for
infinitely wide `slabs' and
systems with periodic boundaries at $i=\pm W$
with $W>M$, where $j=M$ denotes the
maximum height of the packing.

We can now describe the manner in which
the geometric disorder propagates upwards from the base.
For $j\gg1$, the expression (\ref{e:delta_x_ij}) is
dominated by the two sums.
Thus, to good approximation,
the coefficients of both the ${\bf b}^{+}$
and ${\bf b}^{-}$ components of $\delta{\bf x}_{ij}$
are the sum of $j$ independent random variables
$2\delta r_{kl}$\,.
Each such term has a variance $4\sigma^{2}_{\delta r}$,
and so the combined variance is
$\sim4j\sigma^{2}_{\delta r}$.
Therefore the $x$ and $y$--components
of $\delta{\bf x}_{ij}$ will both broaden like
${\mathcal O}(j^{1/2}\sigma_{\delta r})$,
{\em i.e.} the geometric disorder {\em diffuses}
upwards in a normal (Brownian) manner.
This contrasts with the non--Brownian
diffusion of forces described
in Sec.~\ref{s:force_diffusion}.

For frictionless discs, the contact force between
any two contacting beads is
parallel to the line connecting the bead centres,
which can be derived using~(\ref{e:delta_x_ij}).
For instance, if the direction from the centre of bead
$(i,j)$ to $(i+1,j+1)$ is denoted
$\n^{+}+\delta\n_{ij}^{+}$, then

\begin{eqnarray}
2r\delta\n^{+}_{ij}
&=&
(\delta r_{ij}+\delta r_{i+1j+1})
({\bf b}^{+}+{\bf b}^{-}-\n^{+})
\nonumber\\
&+&{\bf b}^{-}
\bigg\{
\delta r_{i+j+2,0}-\delta r_{i+j,0}\nonumber\\
&&+2\sum_{m=1}^{j}
[\delta r_{i+(j-m)+2,m}-\delta r_{i+(j-m),m}]
\bigg\}
\nonumber\\
\label{e:delta_n_plus}
\end{eqnarray}

\noindent{}This obeys
$\delta{\n}_{ij}^{+}\cdot\n^{+}=0$,
ensuring that $\n^{+}+\delta\n_{ij}^{+}$ is a unit vector
to first order in $\delta\n_{ij}^{+}$.
A similar expression applies for contacts from
$(i,j)$ to $(i-1,j+1)$, which have direction
$\n^{-}+\delta\n_{ij}^{-}$ with

\begin{eqnarray}
2r\delta\n^{-}_{ij}
&=&
(\delta r_{ij}+\delta r_{i-1j+1})
({\bf b}^{+}+{\bf b}^{-}-\n^{-})
\nonumber\\
&+&{\bf b}^{+}
\bigg\{
\delta r_{i-j-2,0}-\delta r_{i-j,0}\nonumber\\
&&+2\sum_{m=1}^{j}
[\delta r_{i-(j-m)-2,m}-\delta r_{i-(j-m),m}]
\bigg\}
\nonumber\\
\label{e:delta_n_minus}
\end{eqnarray}

\section{Propagation of forces from a single loaded bead}
\label{s:point_load}

Consider the force propagating from 
a vertical load ${\bf F}^{\rm load}=(0,-f^{\rm load})$
applied to a single bead $(k,l)$.
If the packing were ordered,
{\em i.e.} $\delta r_{ij}\equiv0$ everywhere,
${\bf F}^{\rm load}$ would first split equally
into both of the bead's supporting contacts,
and then propagate unaltered along colinear bonds 
to the fixed base.
With disorder $\delta r_{ij}\neq0$, there are two
differences: not only is ${\bf F}^{\rm load}$
shared unequally by the two contacts
(which now generally have different directions),
but also the propagation to the base is no longer
linear but multiply branched.
A qualitative description of the following analysis
can be found in~\cite{cryst_exp2} for close--packed
3D crystals.

Let $f^{\pm}_{ij}$ denote the magnitude of the
contact force from $(i,j)$ to $(i\pm1,j+1)$,
using the convention that compressive contacts
have positive $f^{\pm}_{ij}$.
The force balance equation for
the loaded bead $(k,l)$ is then

\begin{eqnarray}
{\bf F}^{\rm load}
&+&
(\n^{+}+\delta\n^{+}_{k-1l-1})
f_{k-1l-1}^{+}
\nonumber\\
&+&
(\n^{-}+\delta\n^{-}_{k+1l-1})
f_{k+1l-1}^{-}
=
{\bf 0}
\label{e:force_balance}
\end{eqnarray}

\noindent{}By writing ${\bf F}^{\rm load}$ as
$-(f^{\rm load}/2n_{y})(\n^{+}+\n^{-})$,
the supporting forces can be evaluated to first order
in the $\delta\n^{\pm}$,

\begin{eqnarray}
f_{k-1l-1}^{+}
=
\frac{f^{\rm load}}{2n_{y}}
\left\{
1+
\frac{\phi\n^{-}\cdot\delta\n^{+}_{k-1l-1}
-\n^{+}\cdot\delta\n^{-}_{k+1l-1}}{1-\phi^{2}}
\right\}
\nonumber\\
f_{k+1l-1}^{-}
=
\frac{f^{\rm load}}{2n_{y}}
\left\{
1+
\frac{\phi\n^{+}\cdot\delta\n^{-}_{k+1l-1}
-\n^{-}\cdot\delta\n^{+}_{k-1l-1}}{1-\phi^{2}}
\right\}
\nonumber\\
\label{e:split}
\end{eqnarray}

\noindent{}using $\phi=\n^{+}\cdot\n^{-}$.
The range of allowed $s$
given in Sec.~\ref{s:geometry} means that
$-\frac{1}{2}<\phi<\frac{1}{2}$,
so that $1-\phi^{2}>\frac{3}{4}$
and the denominators in (\ref{e:split})
never diverge.

Once the two contact forces $f_{k\mp1l-1}^{\pm}$
are known, it is necessary to
calculate how they propagate to the fixed base.
This is done by the same force balance equation
as (\ref{e:force_balance}) but with ${\bf F}^{\rm load}$
replaced by the corresponding incoming contact force.
For instance, the force $f^{+}_{rs}$
coming into a bead $(r,s)$ along the
$-(\n^{+}+\delta\n^{+}_{rs})$ direction
is balanced by the two supporting contacts
$f^{\pm}_{r\mp1,s-1}$, which
to order ${\mathcal O}(\delta\n^{\pm})$ are

\begin{eqnarray}
\frac{f^{+}_{r-1s-1}}{f^{+}_{rs}}=&1&+
\frac{\phi}{1-\phi^{2}}
\n^{-}\cdot(\delta\n^{+}_{r-1s-1}-\delta\n^{+}_{rs})
\nonumber\\
\frac{f^{-}_{r+1s-1}}{f^{+}_{rs}}=&&
-\frac{1}{1-\phi^{2}}
\n^{-}\cdot(\delta\n^{+}_{r-1s-1}-\delta\n^{+}_{rs})
\label{e:propagate}
\end{eqnarray}

\noindent{}The expressions for the contributions
from $-f^{-}_{rs}(\n^{-}+\delta\n^{-}_{rs})$
follow from symmetry.
Thus the contribution of each contact force
of the loaded bead $f^{\pm}_{k\mp1,l-1}$ to any
contact $f^{\pm}_{ij}$ in the system, if any,
can be found by summing over all
downward--propagating paths from $(k,l)$ to $(i,j)$, 
and applying (\ref{e:propagate}) for each step.
This process is simplified by the observation that
the ``branching'' force
$f^{-}_{r+1s-1}$ in (\ref{e:propagate})
is ${\mathcal O}(\delta\n^{\pm})$
to leading order, and thus
only paths with 0 or 1 branch point
need to be considered within this first order calculation.

Without loss of generality we now consider the propagation
of the $f_{k-1l-1}^{+}$ forces;
the equivalent $f_{k+1l-1}^{-}$ results follow from symmetry.
Suppose a contact $f_{ij}^{+}$ lower down the packing
obeys $k-i=l-j$, so that it is connected to $f_{k-1l-1}^{+}$
along a straight path with no branches.
Then (\ref{e:propagate}) can be applied iteratively
to give

\begin{equation}
\frac{f_{ij}^{+}}{f_{k-1l-1}^{+}}
=
1
+
\frac{\phi}{1-\phi^{2}}
\n^{-}\cdot
\left(\delta\n_{ij}^{+}
-\delta\n^{+}_{k-1l-1}\right)
\label{e:unidirn}
\end{equation}

\noindent{}Only the $\delta\n^{+}$ at each end
of the path contribute: the intermediate terms
cancel, to first order.

Contacts $f^{-}_{ij}$ with the opposite orientation
may still receive a contribution from $f^{+}_{k-1l-1}$
if a path with precisely one branch point connects the two bonds.
This happens if $n=\frac{1}{2}[(l-j)-(k-i)]$ obeys $1\leq n<l-j$,
in which case the branch point is at bead
$(i-n,j+n)$.
As the branch is ${\mathcal O}(\delta\n^{\pm})$
to leading order, only the zero'th order contributions
from the remaining propagation terms are required,
giving

\begin{equation}
\frac{f^{-}_{ij}}{f^{+}_{k-1l-1}}
=
-
\frac{1}{1-\phi^{2}}
\n^{-}\cdot
\left(\delta\n^{+}_{i-n-1,j+n-1}
-\delta\n^{+}_{i-n,j+n}\right)
\label{e:kink}
\end{equation}

\noindent{}All other $f^{\pm}_{ij}$ receive zero
contribution.
Thus the contribution to any contact force
$f^{\pm}_{ij}$ due to the load ${\bf F}^{\rm load}$
applied to $(k,l)$ can be found by combining
the expressions (\ref{e:split}),
(\ref{e:unidirn}) and (\ref{e:kink}).

\section{Distribution of forces $P(f)$}
\label{s:force_diffusion}

When a specified load is applied to the pile,
the total contribution to any contact force $f^{\pm}_{ij}$
in the packing
can be found by applying the results of Sec.~\ref{s:point_load}
to each loaded bead, and summing.
The distribution of contact forces $P(f)$ can then be
found by either averaging over all contacts with the same height,
or by ensemble averaging over different packings
$\{\delta r_{ij}\}$;
for the loads considered here, which do not vary
with horizontal distance, the final result is the same.
Two alternative loadings are analysed:
a {\em surface} load ${\bf F}^{\rm load}=(0,-f^{\rm load})$
applied to all beads at $j=M$ with all other beads
weightless;
and a {\em bulk} load in which each bead has a weight
$(0,-mg)$.
Only the first case will be treated in detail here;
the results of the qualitatively--similar
bulk loading will be simply stated at the end.

Each contact $f_{ij}^{+}$ will have contributions
from the surface bead $(i+M-j,M)$ according to
the unidirectional propagator (\ref{e:unidirn}),
plus another $M-j-1$ contributions from beads
in the range $(i-M+j+2,M)$ to $(i+M-j-2,M)$ which
branch once according to (\ref{e:kink}).
Including the initial splitting of ${\bf F}^{\rm load}$
at the surface~(\ref{e:split}),
the final expression for $f_{ij}^{+}$ is

\begin{eqnarray}
\lefteqn{
\frac{f_{ij}^{+}}{f^{\rm load}/2n_{y}}
=}
\nonumber\\
&&1+
\frac{1}{1-\phi^{2}}
\left(
\phi\n^{-}\cdot\delta\n^{+}_{ij}
-\n^{+}\cdot\delta\n^{-}_{i+M-j+1,M-1}
\right)
\nonumber\\
&&
-\frac{1}{1-\phi^{2}}
\sum_{n=1}^{M-j-1}
\n^{+}\cdot
\left(
\delta\n^{-}_{i+n+1,j+n-1}
-\delta\n^{-}_{i+n,j+n}
\right)
\label{e:f_total}
\end{eqnarray}

\noindent{}In principle, it is possible to write this expression
in terms of the $\delta r_{ij}$;
however, this soon becomes messy due to the
proliferation of terms.
Instead we now assume that $M-j\gg1$, so that
$(i,j)$ is `sufficiently' below the surface.
In this limit, the second term on the right hand
side of (\ref{e:f_total}) is much smaller than the sum
and can therefore be dropped
(recall that $\delta\n^{\pm}_{ij}\sim{\mathcal O}(j^{1/2})$).
Furthermore, the sum itself will be dominated
by terms with $n={\mathcal O}(M-j)$,
so we need only consider $\delta\n^{-}_{rs}$ with
$s={\mathcal O}(M-j)$.
This means that the sum terms in the expressions
for $\delta\n^{\pm}_{rs}$
(\ref{e:delta_n_plus},\ref{e:delta_n_minus})
will dominate and the remaining terms can be dropped.
Combining these approximations,
$f^{+}_{ij}$ can be written purely in terms of
combinations of $\delta r_{rs}$ like
$2\delta r_{rs}-\delta r_{r-2,s}-\delta r_{r+2,s}=\Delta_{rs}$\,,

\begin{eqnarray}
\frac{f_{ij}^{+}}{f^{\rm load}/2n_{y}}-1
&\approx&
-\frac{1}{r(1-\phi^{2})}
\sum_{n=1}^{M-j}
\sum_{m=1}^{j+n}
\Delta_{i-(j-m),m}
\nonumber\\
&\approx&
-\frac{1}{r(1-\phi^{2})}
\sum_{m=1}^{M}
\sum_{n={\rm max}(1,m-j)}^{M-j}
\Delta_{i-(j-m),m}
\nonumber
\end{eqnarray}
\begin{equation}
\approx
-\frac{1}{r(1-\phi^{2})}
\sum_{m=1}^{M}
\left\{
\begin{array}{r@{\;:\;}l}
M-j & m\leq j \\
M-m & m>j
\end{array}
\right\}
\Delta_{i-(j-m),m}
\nonumber\\
\label{e:Delta}
\end{equation}

\noindent{}Since the mean of each $\Delta_{rs}$ is zero,
we can immediately see that the mean force
$\langle f\rangle=f^{\rm load}/2n_{y}$ for all heights.
Also, since the $\Delta_{i-(j-m),m}$ are independent
for each value of~$m$,
then $f^{+}_{ij}$
is the sum of a number of independent random variables,
each of which has a weighting of no more than
${\mathcal O}(1/M)$ of the total.
Hence the central limit theorem applies, and,
assuming that the variance of bead radii $\sigma^{2}_{\delta r}$
is finite,
the distribution of contact forces must be Gaussian.
The variance of forces $\sigma^{2}_{f}(j)$ depends on the
height $j$ and can be found by standard techniques,
given the weightings of the
$\Delta_{i-(j-m),m}$ shown in~(\ref{e:Delta}).
Noting that the variance of $\Delta_{rs}$ is
$6\sigma^{2}_{\delta r}$,
we finally find that

\begin{equation}
\frac{\sigma^{2}_{f}(z)}{\langle f\rangle^{2}}
\sim
\frac{2M^{3}}{(1-\phi^{2})^{2}}
(1-z)^{2}(2z+1)
\left(
\frac{\sigma_{\delta r}}{r}
\right)^{2}
\label{e:surface_diffusion}
\end{equation}

\begin{figure}
\centerline{\psfig{file=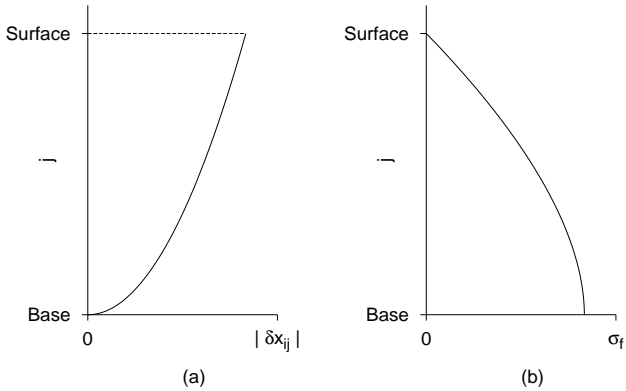,width=9cm}}
\caption{Graphical representation of
(a)~the broadening of the geometric disorder with
height $j$, and
(b)~the broadening of the force distribution with
distance from the surface,
as given by the surface--loaded
solution~(\ref{e:surface_diffusion}).
}
\label{f:schematic}
\end{figure}

\begin{figure}
\centerline{\psfig{file=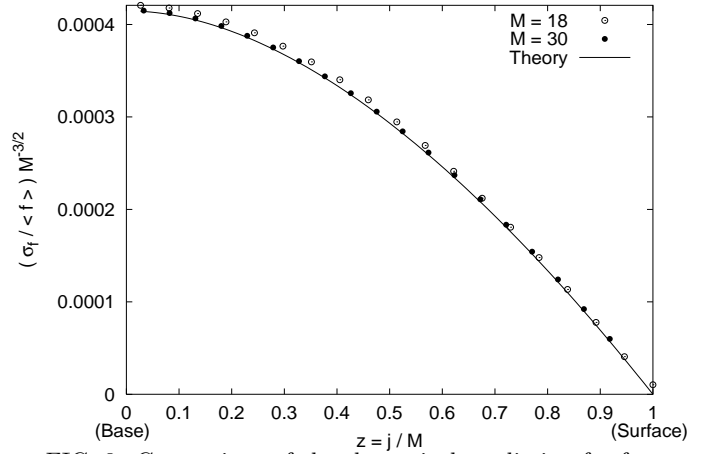,width=9cm}} 
\caption{Comparison of the theoretical prediction
for force diffusion for a surface--loaded
system~(\ref{e:surface_diffusion}) with
corresponding simulation data.
The normalised standard deviation of forces
$\sigma_{f}/\langle f\rangle$ has been
scaled by $M^{3/2}$ so that,
according to~(\ref{e:surface_diffusion}),
the curve will be $M$--independent when plotted
against the relative height $z=j/M$.
The two pile heights used in the simulations
are given in the key, and appear to confirm this
prediction.
The parameters for both theory and simulations are
$(\sigma_{\delta r}/r)^{2}=10^{-6}/12$,
$r=1$ and $s=3$.
}
\label{f:sims}
\end{figure}

\noindent{}where $z=j/M$.
A graphical representation of this solution is
given in Fig.~\ref{f:schematic}.
This should be compared with
the Brownian expectation $\sigma_{j}^{2}\sim1-z$,
as seen in {\em e.g.} the
$q$--model~\cite{jaco_1,jaco_2}.
Not only is (\ref{e:surface_diffusion})
superdiffusive near the surface,
$\sigma^{2}_{j}\sim(1-z)^{2}$,
the prefactor depends on $M$,
{\em i.e.} the depth of the pile is
`felt' at the surface, no matter how large $M$ is.
This is not unexpected, since the degree of
geometric disorder at the surface always depends on
$M$ in these near--crystalline packings;
for more disordered systems, it
would most likely saturate at some finite height and
the $M$--dependence near the surface would vanish.
Also note that the expression (\ref{e:surface_diffusion})
becomes subdiffusive nearer the base,
and ceases to broaden precisely at $j=0$.
This is because the geometric disorder vanishes at the
base, so that the forces no longer diffuse.
To check the assumptions leading to
(\ref{e:surface_diffusion}), we compare it to
data obtained from the adaptive network
algorithm~\cite{adap_net} in Fig.~\ref{f:sims}.
The agreement is good, and appears to improve
as $M$ is increased.

For a bulk loaded system, the above calculations
can be repeated in a similar manner.
The main difference is that the mean force
now varies with depth,
$\langle f\rangle_{z}=Mmg(1-z)/2n_{y}$.
Nonetheless, the variance normalised by the
mean force for each depth takes a similar form,

\begin{equation}
\frac{\sigma^{2}_{f}(z)}{\langle f\rangle^{2}_{z}}
\sim
\frac{3M^{3}}{10(1-\phi^{2})^{2}}
(1-z)^{2}(4z+1)
\left(
\frac{\sigma_{\delta r}}{r}
\right)^{2}
\label{e:bulk_diffusion}
\end{equation}

\noindent{}The right hand side of this expression
reaches a maximum at finite
height $j^{*}=M/6$, and then {\em narrows} slightly
towards to the base.
This is because the diffusion of forces decreases
near the base,
as mentioned above,
but the mean force
(to which the distribution is normalised) increases
linearly with $1-z$, which acts to narrow the distribution.
If the distribution was {\em not} normalised to
$\langle f\rangle_{z}$, then it would
broaden monotonically and again cease to broaden
just at the base,
for the same reasons as the surface loading case.

\subsection{Breakdown of the compressive bond assumption}
\label{s:M}

We can now discuss the range of validity
of the assumptions made earlier.
The magnitude of the displaced bead centres
$\delta{\bf x}_{ij}$ typically vary as
${\mathcal O}(M^{1/2}\sigma_{\delta r})$,
which will be similar to the bead radii $r$,
and hence violate the assumed contact topology,
for piles of height
${\mathcal O}[(\sigma_{\delta r}/r)^{-2}]$
and greater
(this assumes that the gaps between beads in the
ordered packing are also ${\mathcal O}(r)$).
However, the assumption of non--tensile contacts
will be violated much sooner.
For both surface and bulk loading,
$\sigma_{f}$ averaged
over the system has the same general form

\begin{equation}
\frac{\sigma_{f}}{\langle f\rangle}
\propto
M^{3/2}
\frac{\sigma_{\delta r}}{r}
\label{e:general_form}
\end{equation}

\noindent{}Thus
a finite fraction of forces will become negative
for packings of height
$M^{*}={\mathcal O}[(\sigma_{\delta r}/r)^{-2/3}]$,
and the above analysis only holds for $M\ll M^{*}$.
Note that this is significantly smaller than if the
forces diffusion in a Brownian manner
$\sim M^{1/2}$.
Strictly speaking, we should really require that
a vanishing {\em number} of
contact forces are negative rather than
a vanishing fraction,
which would give a much lower $M^{*}$.
However, this is most likely too restrictive, unless
it could be shown that {\em e.g.} a single negative force
initiates a cascade of rearrangements that alters
the forces throughout a finite fraction of the system.


\section{Response Green's function}
\label{s:greens}

The various phenomenological theories for granular
stress propagation differ most markedly in the response
Green's function to an isolated
load~\cite{proc,jpb_proceedings,doubleY3,geng}.
This can be calculated for our system as follows:
apply a load $(0,-f^{\rm load})$ to bead $(k,l)$,
and measure the vertical component $p$ of the induced
force at a contact $f^{\pm}_{ij}$.
This can be found for a given packing $\{\delta r_{ij}\}$
using the results given above.
Now {\em ensemble} average over different
packing realisations, {\em i.e.} uncorrelated sets of
$\{\delta r_{ij}\}$ with the same variance $\sigma^{2}_{\delta r}$.
Since this is a first--order calculation, the mean response
will be as the ordered case, so that
$\langle p\rangle_{\rm ensemble}$ will be zero
except along the cone propagating downwards from $(k,l)$.
However, although the response outside this cone is
strictly zero, the response inside the cone only vanishes
after averaging: it will be positive or negative for each
particular packing.
The variance of the distribution of $p$
between different realisations
can be calculated using a similar procedure to
that given in Sec.~\ref{s:force_diffusion},

\begin{equation}
\frac{\sigma^{2}_{\rm ensemble}(p)}{(f^{\rm load})^{2}}
\sim
\frac{3}{2(1-\phi^{2})^{2}}
(l+j)
\left(
\frac{\sigma_{\delta r}}{r}
\right)^{2}
\label{e:greens1}
\end{equation}

\begin{figure}
\centerline{\psfig{file=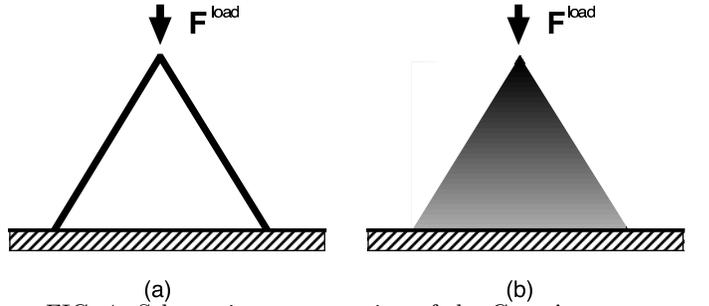,width=9cm}}
\caption{Schematic representation of the Green's response
solution (\ref{e:greens1}) to a
vertical load ${\bf F}^{\rm load}$.
$(a)$~The mean response is identical to the corresponding
ordered system,
in which forces propagate along lines of colinear
bonds forming a `cone.'
$(b)$~The magnitude of the fluctuations
under ensemble--averaging, 
with darker grey corresponding to a greater variance.
The variance is zero outside the cone,
depends only on height inside the cone,
and is twice as large near the apex as near the base
(fluctuations on the cone itself are not shown).
}
\label{f:greens}
\end{figure}

\noindent{}A schematic representation of this solution
is given in Fig.~\ref{f:greens}.
Interestingly, this does {\em not} decay away from the
edges of the cone,
as predicted by phenomenological approaches,
and in fact is completely independent
of the horizontal position, depending only on the
combined heights $l$ and~$j$.

The counterpart to ensemble averaging is
coarse graining, {\em i.e.} applying a uniform
pressure to a group of $N$ beads, and jointly taking
$N$ and $j$ to infinity in a suitable manner
to give a continuum result.
However, for our packings this will just give
a zero response everywhere except along the
edge of the cone.
This is because we have only expanded around
the ordered solution to first order, which will always
give the zero--th order, ordered result after any
form of averaging.
Extending these calculations to second order will
introduce quadratic terms that do not vanish under
averaging, and thus would allow a direct comparison
between different types of averaging.

\section{Summary and discussion}
\label{s:disc}

In summary, we have obtained analytical expressions
for contact forces in isostatic packings of
frictionless discs in the nearly--ordered limit.
These explicitly demonstrate the anomalous broadening
of the distribution of forces $P(f)$ with increasing depth,
and the unusual response Green's function, whose
magnitude of fluctuations
under ensemble averaging depends only
on the vertical coordinate.
We see no reason in principle why these findings could
not be tested experimentally, as long as a sufficiently
monodisperse sample of smooth beads could be found.
Indeed, this may be the main problem:
glass bead experiments with a low--polydispersity
of sizes placed in close--packed crystalline configurations
have shown that $P(f)$ already has a broad exponential $P(f)$
at the base~\cite{cryst_exp2,cryst_exp1},
as opposed to the Gaussian form predicted here.
This suggests that even these carefully controlled
experiments are already too disordered to see the
first--order results predicted here.

It is hoped that the calculations presented in this
paper may lay the foundations for a more complete,
analytically--tractable theory of bead packings that
would complement current approaches.
Thus it is worth discussing in what ways these calculations
may be extended to more realistic situations.
There should be little problem in going to 3 dimensions,
or for considering loads at an angle to the vertical,
which would allow the piles to be sheared.
Perhaps more interesting would be to expand about
the ordered solution to second order in the $\delta r_{ij}$,
rather than just to first order as considered here.
This would allow the mean force to deviate from
the crystalline solution for the first time,
and hence allow the nature of force propagation
in granular packings to be addressed from within
an analytical framework; for instance, it would be
possible to see if proposed relationships between
the components of the stress tensor are valid~\cite{proc,alexei1}.
It would also show if reaching the continuum limit by
ensemble averaging differs from coarse graining,
as already discussed in Sec.~\ref{s:greens}.

However, extending this analysis
to frictional beads is likely to be more troublesome.
The marginal rigidity state for beads with
infinite friction has $Z=3$ in 2 dimensions, so that
sequentially deposited beads will make either
1 or 2 contacts on the surface -- in the case of
1 contact, it is not possible to determine the
rest position of the incoming bead. Thus groups
of beads must be added simultaneously, so that
the final geometry will depend on incoming bead
trajectories and their material properties.
Also, with $Z=3$ it is not possible to uniquely decompose
the load applied to any given bead along paths
connecting it to the base, for either
the normal or tangential contact forces.
Instead loops, and also paths leading to the surface,
will inevitably arise.
This non--unique decomposition will also be a problem for
frictionless non--spherical beads, for which $Z=6$ in
2 dimensions.
It is not clear how these problems may be surmounted
without resorting to {\em ad hoc} assumptions or
simplifications.


\section*{Acknowledgments}

The author would like to thank Mike Cates and Jacco
Snoeijer for correspondence relating to this work.
This research was partly funded through a
European Community Marie Curie Fellowship.


\end{multicols}


\begin{references}
\bibitem{formal} Formally: start from the ordered packing,
for which every contact force is known to be finite
and compressive.
Now increase the polydispersity of bead radii
continuously from zero.
Since the contact forces will also vary continuously
(for systems of finite size),
they must remain compressive for some finite range
of polydispersity.
\end{references}
\end{document}